\documentclass[aps,prl,twocolumn,groupedaddress,floatfix]{revtex4}

\usepackage{epsfig}
\usepackage{graphicx}
\usepackage{float}
\usepackage{dcolumn}
\usepackage{amsmath}
\usepackage{amssymb}
\input{epsf}

\ifx\pdftexversion\undefined
  \usepackage{graphicx}
\else
  \usepackage[pdftex]{graphicx}
\fi

\newcommand{\ket}[1]{\left| {#1} \right\rangle}

\def\D{\Delta}
\def\w{\omega}
\def\a{\alpha}

\def\k{\kappa}
\def\W{\Omega}
\def\g{\gamma}
\def\G{\Gamma}

\def\be{\begin{equation}}
\def\ee{\end{equation}}
\def\bea{\begin{eqnarray}}
\def\eea{\end{eqnarray}}

\def\>{\rangle}
\def\<{\langle}

\begin{document}

\title{Coherent single-photon generation and trapping with practical cavity QED systems}

\author{David Fattal$^{1,2}$, Raymond G.\ Beausoleil$^{1}$, and Yoshihisa Yamamoto$^{2}$}

\affiliation{$^1$Hewlett-Packard Laboratories, 1501 Page Mill Road,
Palo Alto, CA 94304}

\affiliation{$^2$Quantum Entanglement Project, SORST, JST, Ginzton
Laboratory, Stanford University, Stanford CA 94305}

\email{david.fattal@hp.com}


\begin{abstract}
We study analytically the dynamics of cavity QED nodes in a
practical quantum network. Given a single 3-level $\Lambda$-type
atom or quantum dot coupled to a micro-cavity, we derive several
necessary and sufficient criteria for the coherent trapping and
generation of a single photon pulse with a given waveform to be
realizable. We prove that these processes can be performed with
practical hardware --- such as cavity QED systems which are
operating deep in the weak coupling regime --- given a set of
restrictions on the single-photon pulse envelope. We
systematically study the effects of spontaneous emission and
spurious cavity decay on the transfer efficiency, including the
case where more than one excited state participates in the
dynamics. This work should open the way to very efficient
optimizations of the operation of quantum networks.
\end{abstract}



\maketitle


Photons, the elementary constituents of light, are fast and robust
carriers of quantum information. Recently, techniques have been
found to reversibly produce or trap photons one by one in matter.
This inter-conversion capability potentially allows quantum
information processing using the best of two worlds: the fast and
reliable transport of quantum information via photons, and its
storage and processing in matter where interactions between qubits
can be made strong. Optical cavities with the ability to
concentrate the electro-magnetic field in small regions of space
provide a natural interface for photonic and matter qubits. They
constitute a key element of quantum networks\cite{ref:Qnet} in
which cavity ``nodes'' communicate coherently via photonic
channels\cite{ref:2}. The theory of photon absorption and trapping
was partially worked out in \cite{ref:Sham_05}, extending the
initial solution given in \cite{ref:Qnet}. This approach relies on
a carefully designed classical ``control'' pulse driving a 3-level
quantum system in a $\Lambda$ configuration, and numerically shows
that (contrary to widely accepted belief) the photon transfer can
sometimes be realized in a non-adiabatic (i.e., rapid) way. Until
then, both theory\cite{ref:8} and experiments\cite{ref:3, ref:4,
ref:5, ref:6,ref:7} aiming at coherent photon emission from a
driven cavity-QED system were based on an adiabatic technique
called Stimulated Raman Adiabatic Passage (STIRAP), requiring a
slowly varying control pulse. Another widely accepted requirement
for coherent photon transfer is the one of strong coupling between
the cavity and the atomic system. This assumption was also
challenged numerically in Ref.~\cite{ref:Imamoglu_04}, where it
was shown that (adiabatic) coherent photon generation with high
efficiency and indistinguishability was possible in the weak
coupling regime. However, there remain a number of open questions
regarding this technique: Exactly what kind of single photons can
be generated and/or trapped? How fast can their envelopes vary?
How far detuned from the cavity resonance can they be, and what
Raman detunings can be tolerated in connected cavity nodes? How
weak a coupling between cavity and atom can be tolerated, and at
what expense in performance? Finally, how sensitive is the
transfer technique to characteristics of non-ideal systems with
spontaneous emission, spurious cavity decay, or where several
excited states contribute to the dynamics?

\begin{figure}[htbp]
    \epsfig{file=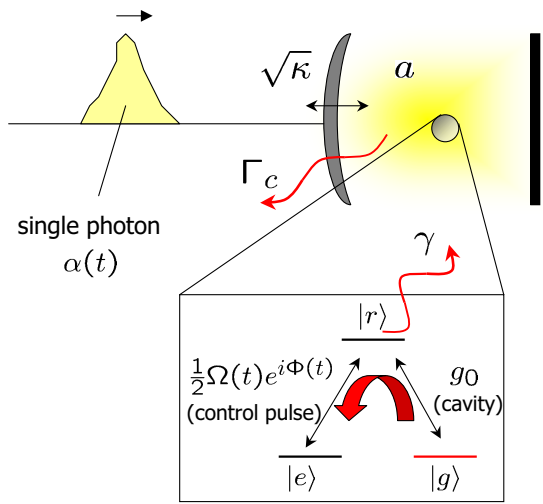, width=3in}
    \caption{Illustration of a node, a 3-level $\Lambda$ system in a single mode cavity.
    The red arrow represents the photon trapping process, where a photon from the waveguide
    is absorbed by the node and induces the flip $\ket{g} \rightarrow
    \ket{e}$, if the correct control pulse $\W(t)$ is applied
    to the $e-r$ transition. The loss mechanisms are represented by the red wavy arrows.}
    \label{fig:scheme}
\end{figure}

In this paper, we derive analytical formulas that answer all of
these questions, and provide a solid basis for the optimization of
quantum networks. When losses are neglected, we show that a single
criterion --- given by Eq.~(\ref{crit}) below --- determines the
existence condition for a control pulse achieving coherent photon
transfer in a 3-level cavity-QED system. This criterion tells us
which complex envelope functions $\a(t)$ of a single photon pulse
are eligible for generation and/or trapping when the
characteristics of the cavity-QED system are known. Interestingly,
this criterion can be satisfied in the non-adiabatic or in the
weak-coupling regime, in the presence of photon-cavity detuning,
and for arbitrary Raman detuning $\Delta$ (see
Fig.~\ref{fig:scheme}), suggesting that quantum networks could be
operated even with highly deficient and heterogeneous nodes. When
losses are included, we find another criterion for the existence
of a control pulse that leaves the node and the waveguide
unentangled after transfer, and we provide an analytical
expression for the transfer efficiency. We show that very
efficient transfer can be realized in the weak coupling regime
provided the Purcell factor is sufficiently high, regardless of
whether strong coupling is precluded by a low cavity $Q$ (as in
solid-state systems) or by a large spontaneous emission rate (as
in trapped ions and atoms) when compared to the vacuum Rabi
frequency. We generalize the loss analysis to the case where other
excited states participate in the dynamics of the transfer.


Consider the system shown in Fig.~\ref{fig:scheme}, a 3-level
$\Lambda$ atom with ground states $\ket{g}$ and $\ket{e}$ and
excited state $\ket{r}$ encapsulated within a single-mode cavity.
The cavity vacuum field and coherent control field couple levels
$g-r$ and $e-r$ respectively, with Rabi frequency $g_0$ and
$\frac{1}{2} \W(t)$, and common (Raman) detuning $\Delta$. The
cavity mode with frequency $\w_c$ leaks into a single waveguide
mode, with a rate $\k$. The first important result of the paper is
the following: in the absence of loss, given an incident single
photon pulse with complex amplitude $\alpha(t)e^{-i \w_c t}$
satisfying at all times the criterion
\be \int_{-\infty}^t |\a(s)|^2 ds - \frac{|\a(t)|^2}{\k} -
\frac{1}{\k g_0^2}|\dot{\a}(t) - \frac{\k}{2} \a(t)|^2 > 0 ,
\label{crit}\ee
there exists a unique control pulse $\Omega(t) = |\Omega(t)| e^{i
\Phi(t)}$ that achieves deterministic trapping of the photon,
driving the system from initial state $\ket{g}$ to final state
$\ket{e}$. Before going into the details of the proof, we study
the restrictions imposed by (\ref{crit}). First, note that it is
independent of the Raman detuning $\Delta$, so that changing
$\Delta$ need not cause any failure as long as the control pulse
is changed accordingly. In practice, the intensity and chirp of
the control pulse need to increase with $\D$, setting experimental
limitations on the amount of detuning. Criterion~(\ref{crit}) is
easier to satisfy if $g_0$ is large, and if the photon pulse is
slow. In most cases of interest the bandwidth of the photon pulse
should be always smaller than both $\k$ and $g_0^2/\k$ --- bearing
in mind that these are estimates that can be violated for short
time durations. In the strong coupling regime, this means that the
maximum bandwidth of the photon pulse is $\k$, irrespective of the
cavity coupling. In the weak coupling regime, photon transfer can
still occur with efficiencies approaching unity if the photon is
slowly varying enough, with a bandwidth not greater than
$g_0^2\k$. Finally, we note that the scheme can accommodate a
photon-cavity detuning on the order of $g_0$; larger detunings
would increase the negative contribution of $\dot{\a}(t)$ to
(\ref{crit}), making the criterion difficult to satisfy in
practice.

The second important set of results concern the transfer
efficiency in the presence of spontaneous emission from level
$\ket{r}$ (with rate $\g$) and spurious cavity losses (with rate
$\G_c$). These dissipative processes are not time-reversal
symmetric, so that the photon generation problem has to be handled
differently from the photon trapping problem. In both cases, we
find a control pulse that achieves the best transfer efficiency
possible while leaving the waveguide and the cavity QED node in a
separable (unentangled) final state. Technically, such a pulse may
not give the absolute highest efficiency for the transfer, but has
the key feature that it will minimize the propagation of errors in
the network. In the trapping case, the control pulse satisfying
the disentanglement condition exists for a certain class of
envelope function $\a$ obeying a criterion similar to
(\ref{crit}), but taking into account the modification of coherent
dynamics introduced by $\g$ and $\G_c$. We will show that the
transfer efficiency is given by
\be \eta_\text{trap} = \left(1 - \frac{\G_c}{\k}\right) \left(1 -
\frac{\g(\k-\G_c)}{4g_0^2} \right) - \frac{\g}{\k g_0^2} \int
|\dot{\a}|^2 \label{eta_trap}\ee
Whether the transfer succeeds or not, the photon has been
perfectly removed from the waveguide.  In the photon generation
case, the eligible photon waveforms are also determined by a
modified criterion, and the transfer efficiency is:
\be \eta_\text{gen} = \left[\left(1+\frac{\G_c}{\k} \right)
\left(1+\frac{\g(\k+\G_c)}{4g_0^2} \right) + \frac{\g}{\k g_0^2}
\int |\dot{\a}|^2\right]^{-1} \label{eta_gene}\ee
These relations indicate that efficient transfer in the presence
of losses is possible if the cavity-waveguide interface is well
designed ($\G_c \ll \k$) and if the quantity $4g_0^2/\k\g$ (equal
to the Purcell factor in the weak coupling regime) is
significantly greater than one.

We now prove the above claims and exhibit the corresponding
control pulses.  We make the simplifying assumptions that there is
no additional dephasing of the $e-r$ and $g-r$ transition beyond
$\g/2$, and we assume that level $\ket{r}$ decays mostly in
$\ket{g}$ or outside of the system. Under these conditions it is
possible to treat the state of the node as a pure state, defined
as
\be \Psi(t) = g(t) \ket{g,1} + r(t) \ket{r,0} + e(t) \ket{e,0} ,
\ee
where the notation $\ket{X,n}$ stands for the node in level $X$
with $n$ photons in the cavity. Note that $||\Psi(t)||^2$ is the
probability of finding the excitation in the node (either in the
cavity or in the atom) at time $t$. If these conditions do not
hold, the state of the system becomes mixed and a density matrix
approach is required. In the quantum jump picture
\cite{ref:Dalibard_quantumjumps}, the system undergoes a coherent
evolution described by $\Psi(t)$, but at each time step it has
some probability of collapsing to state $\ket{e,0}$ (due to decay
to level $e$) or of undergoing the change $r(t) \rightarrow -r(t)$
(due to dephasing). In this case equations (\ref{eta_trap}) and
(\ref{eta_gene}) provide a lower bound for the average fidelity of
the transfer.

Using the standard cavity input-output relations \cite{Walls} and
the critical assumption that no more than one photon can be
present in the leakage mode of the cavity, we can show that in the
rotating wave approximation, the evolution of the waveguide state
and of $\Psi(t)$ are given by:
\bea
\a_{out} & = &  \sqrt{\k} g - \a_{in} \label{c1}\\
\dot{g} & = & -ig_0^{\ast} \,r -\frac{\kappa + \G_c}{2}g + \sqrt{\kappa} \a_{in} \label{c2}\\
\dot{r} & = & -(\frac{\g}{2}+i\Delta) r -ig_0\,g -i\frac{\Omega}{2}\,e \label{c3}\\
\dot{e} & = & -i\frac{\Omega^{\ast}}{2}\,r \label{c4} \eea
To study the trapping of a photon $\a(t)$, we set $\a_{in} = \a$
and impose the condition $\a_{out}=0$. We find that
\be \frac{\W}{2} = \frac{i}{e} \left[ \dot{r} +
\left(\frac{\g}{2}+i\Delta\right) r + ig_0 g \right] ,
\label{om}\ee
where the necessary and sufficient condition for the existence of
$\W(t)$ is that $|e| > 0$ on all finite time intervals
\cite{ref:laudenbach}. In the above expression, $g = \a/\sqrt{\k}$,
$r = i (\dot{g} - \frac{\k-\G_c}{2}g)/g_0^{\ast}$, and $e$ is given
by
%
%
\bea |e|^2 & = & \left(1 - \frac{\G_c}{\k}\right) \left(1 -
\frac{\g(\k-\G_c)}{4g_0^2} \right) \int^{t} |\a|^2 - \frac{\g}{\k
g_0^2}  \int^{t} |\dot{\a}|^2  \nonumber\\
& & - \frac{|\a|^2}{\k}\left[1 - \frac{\g(\k-\G_c)}{2g_0^2}
\right] -
\frac{|\dot{\a}-\frac{(\k-\G_c)}{2}\a|^2}{\k g_0^2}    \label{E_trap}\\
\dot{\Phi}_e & = & \frac{|r|^2 (\dot{\Phi}_r + \D) - |g|^2
\dot{\Phi}_g }{|e|^2} . \eea
When the right hand side of Eq.~(\ref{E_trap}) is strictly
positive at all times, the control pulse exists, and the transfer
efficiency is given by $|e|^2(t \rightarrow \infty) =
\eta_\text{trap}$. When trapping occurs, the quantum mechanical
amplitude $-\alpha(t)$ associated with the photon being reflected
by the front mirror of the cavity exactly cancels the amplitude
$\sqrt{\k}g(t)$ of the photon being absorbed in the node and
re-emitted from it. This \textit{destructive interference} can be
viewed as an impedance-matching requirement between the incoming
waveguide and the receiving node that has to be insured by a
carefully designed control pulse. Note that for $\G_c > \k$, we
can never find a control pulse that leaves the node and the
waveguide unentangled, irrespective of the photon waveform.

In the case of photon generation, we assume that $\a_{in} = 0$ and
we impose $\a_{out} = \sqrt{\eta}\, \a$, where $\a$ is the desired
photon envelope normalized to 1, and $\eta$, the generation
efficiency, will have to be found self-consistently. The proper
control pulse is still given by (\ref{om}) in this case, but with
$g = \sqrt{\eta \a/\k}$, $r = i (\dot{g} +
\frac{\k+\G_c}{2}g)/g_0^{\ast}$, and
%
%
\bea \frac{1-|e|^2}{\eta} & = &  \left(1+\frac{\G_c}{\k} \right)
\left(1+\frac{\g(\k+\G_c)}{4g_0^2} \right) \int^{t} |\a|^2 +
\frac{\g}{\k g_0^2} \int^{t} |\dot{\a}|^2
 \nonumber\\
 &  & + \frac{|\a|^2}{\k}\left[1 + \frac{\g(\k+\G_c)}{2g_0^2} \right] +
\frac{|\dot{\a}+\frac{(\k+\G_c)}{2}\a|^2}{\k g_0^2}  \label{E_gene}\\
\dot{\Phi}_e & = & \frac{|r|^2 (\dot{\Phi}_r + \D) - |g|^2
\dot{\Phi}_g }{|e|^2} . \eea
The node will be disentangled from the waveguide if and only if
$e(t \rightarrow \infty) =0$, that is $\eta = \eta_\text{gen}$.
The corresponding control pulse then exists if and only if the
right hand side of eq. (\ref{E_gene}) stays strictly positive at
all
times.\\

It is worth noting that in the case of zero Raman detuning and
when the photon pulse itself has no chirp (i.e. when $\a(t)$ can
be taken real), then $\Phi(t)$ can be taken constant: no chirp is
needed on the control pulse, a potentially desired feature for
experiments. If in addition the photon pulse is slow (adiabatic
regime) and the cavity coupling is assumed strong, as in many
photon generation experiments \cite{ref:3, ref:4, ref:5,
ref:6,ref:7}, we can obtain a particularly simple relation between
control and photon pulse:
\be \left| \frac{\Omega(t)}{2} \right|^2 \sim
\frac{g_0^2|\a(t)|^2}{\int_{-\infty}^t|\a(s)|^2\, ds} ,
\label{adiab}\ee
with the dual relation
\be |\a(t)|^2 =
\frac{\kappa}{g_0^2}\left|\frac{\Omega(t)}{2}\right|^2 \,
e^{\frac{\kappa}{2g_0^2} \int_t^{+\infty}\,|\frac{\Omega(s)}{2}|^2
\, ds} . \ee
These expressions explain the experimental observations that a
photon emitted with the STIRAP technique will ``follow'' the
control pulse with some retardance due to the finite value of
$\k$.

We now study the effect of having $N$ excited levels $\ket{r_k}$
contributing to the transfer dynamics. We denote as $g_k$ and
$\W_k$ respectively the couplings of level $\ket{r_k}$ to level
$\ket{g}$ and level $\ket{e}$, and we denote as $\D_k$ and $\g_k$
the corresponding Raman detuning and spontaneous decay rate from
level $r_k$. For simplicity, we gather the quantities $r_k(t)$,
$g_k$ and $\W_k(t)$ into size $N$ vectors $R(t)$, $G$ and $\W(t)\,
V$, with $V$ a vector of unit length. We also define a complex
detuning matrix $\D = \text{diag}(\D_k-i \g_k)$. Using the
notation $y \equiv \dot{g} + (\k+\G_c) g/2 - \sqrt{\k} \a_{in}$,
the evolution of the system is given by
\bea y(t) & = & -i G^{\dagger}\cdot R(t)\\
\dot{R}(t) & = & -i \D \cdot R(t) -i g(t)\,G - i
\frac{\W(t)}{2} e(t) V \label{R_eq}\\
\dot{e}(t) & = & -i \frac{\W^{\ast}(t)}{2} V^{\dagger} \cdot R(t)
\eea
The condition for proper trapping is that $y(t) = \dot{g} -
(\k-\G_c) g/2$. To find the correct control pulse, we split the
vector space into the dimension 1 subspace subtended by vector $V$
and its orthogonal, with corresponding notation $\|$ and $\bot$.
With these notations, the above set of equations can be rewritten
as~:
\bea y & = & -i G_{\|}^{\ast}R_{\|} -i G_{\bot}^{\dagger}\cdot R_{\bot}\\
\dot{R}_{\|} & = & -i \D_{\|}^{\|} R_{\|} -i \D_{\bot}^{\|}\cdot R_{\bot}(t)  -i g\, G_{\|} - i \frac{\W}{2} e\\
\dot{R}_{\bot} & = & -i \D_{\|}^{\bot} R_{\|} -i
\D_{\bot}^{\bot}\cdot R_{\bot} - i g\,
G_{\bot}\\
\dot{e} & = & -i \frac{\W^{\ast}}{2} R_{\|} \eea
When it exists, the correct control pulse envelope function is
given by
\be \frac{\W}{2} = \frac{i}{e} \left(\dot{R}_{\|} +i \D_{\|}^{\|}
R_{\|} +i \D_{\bot}^{\|}\cdot R_{\bot}  +i g\, G_{\|} \right) ,
\ee
where the components of $R$ are given by
\bea R_{\|} & = & \frac{i}{G_{\|}^{\ast}}\left(y + i
G_{\bot}^{\dagger} R_{\bot} \right)\\
R_{\bot} & = & -i e^{-iMt} \int^t e^{iMs} \left(G_{\bot} g(s) +i
\frac{\D_{\|}^{\bot}}{G_{\|}^{\ast}} y(s) \right)  \\
M & = & \D_{\bot}^{\bot} - \frac{\D_{\|}^{\bot}
G_{\bot}^{\dagger}}{G_{\|}^{\ast}} \eea
and the amplitude $e$ is given by
\bea |e|^2 & = & (1-\frac{\G_c}{\k})\int^t |\a|^2 + 2Im(\int^t
R^{\dagger}\D R)\nonumber\\
& &  - \frac{|\a|^2}{\k} - R^{\dagger}R\\
\dot{\Phi}_e & = & \frac{|R_{\|}|^2
\left[Re(\D_{\|}^{\|})+\dot{\Phi}_{R_{\|}} \right] +
Re\left[(\D_{\bot}^{\|}R_{\bot}+g
G_{\|})R_{\|}^{\ast}\right]}{|e|^2} \eea
where again the expression for $|e|^2$ has to be strictly positive
for all times. If $M$ has a purely real eigenvalue, the control
pulse must be turned on for an infinite amount of time to prevent
re-emission of the photon in the waveguide. Even then, a fraction
of the population will be trapped in a ``dark'' state in the
excited states manifold. In general this situation will not happen
due to the spontaneous decay. The additional levels will only
further reduce the trapping efficiency to a value of
\be \eta_\text{trap}^{(N)} = 1 - \frac{\G_c}{\k} + 2 \int
R^{\dagger}\, \text{Im}(\D)\, R . \ee
The same formula hold for photon generation provided we use $y(t)
= \dot{g} + (\k + \G_c) g/2$ and $g =
\sqrt{\eta_\text{gen}^{(N)}/\k}\, \a$, with
\be \eta_\text{gen}^{(N)} = \left[1 + \frac{\G_c}{\k} + 2 \int
R^{\dagger}\, \text{Im}(\D)\, R \right]^{-1} .\ee
%
%
%
Note that the extra absorption caused by the added excited levels
increases when $\a$ has large Fourier components at frequencies
corresponding to the real part of the eigenvalues of $M$. It could
be avoided by a clever design the photon envelope.\\

To summarize, we have derived a series of analytical formulas that
clearly delimit the range of operation of cavity QED nodes for
quantum networks. With these formulas in hand, we are able to prove
that nodes can operate even deeply in the weak coupling regime, at
the expense of slowing down the information transfer. Nodes can be
operated with an arbitrary atom-cavity detuning $\D$ as long as we
have the experimental ability to generate a compensating chirp in
the control pulses. A significant amount of detuning between cavity
photon and cavity resonance can also be tolerated when the vacuum
Rabi frequency $g_0$ is large. This ability will be key to the
operation of heterogeneous cavity QED networks, and suggests that
nodes featuring the highest coupling $g_0$ will be central to the
defect-tolerant operation of the network. For non-ideal systems, we
derived an analytical expression of the transfer efficiencies was
derived. Spontaneous emission from the excited state causes loss by
an amount that is inversely proportional to the Purcell factor
(which can be high even in the weak coupling regime). Spurious
cavity decay starts to cause prohibitive losses when it becomes
comparable to the cavity-waveguide coupling rate. When several
excited levels participate in the node dynamics, additional losses
occur as some energy becomes irremediably trapped in (and radiated
from) a sub-manifold that is not accessible to the control pulse.

The authors acknowledge Charles Santori for his scrupulous
examination of the manuscript. This work was supported in part by
JST SORST, MURI Grant No. ARMY, DAAD19-03-1-0199 and NTT-BRL.





\end{document}